\documentclass{llncs}[13pt]
\usepackage{mathrsfs}
\usepackage[utf8]{inputenc}
\usepackage[english]{babel}
\usepackage{multicol}
\usepackage{pdfpages}
\usepackage{wrapfig}
\usepackage{color}
\usepackage{tabularx}
\usepackage{caption}
\usepackage{bm}
\usepackage{mathrsfs}
\usepackage{amsmath}
\usepackage{enumitem}
\usepackage{pdfpages}
\usepackage{float}
\usepackage{graphicx}
\usepackage{csquotes}
\usepackage{hyperref}
\usepackage{amsmath}
\usepackage{multirow}
\usepackage[margin=1.40in]{geometry}

\newtheorem{theo}{Theorem}

\newtheorem{lem}{Lemma}
\newtheorem{prop}{Proposition}

\numberwithin{equation}{section}
\numberwithin{theorem}{section}
\numberwithin{cor}{section}
\numberwithin{lemma}{section}
\numberwithin{proposition}{section}
\numberwithin{cor}{section}
\numberwithin{eg}{section}
\numberwithin{examp}{section}

\newcommand{\T}{\mathbf{T}}

\newcommand{\bbeta}{\bm{\beta}}

\begin{document}

\title{Distributional Consistency of Lasso by Perturbation Bootstrap}
\titlerunning{Hamiltonian Mechanics}  
%
\author{Debraj Das \and
S. N. Lahiri
}
\authorrunning{Debraj Das} 

\tocauthor{Ivar Ekeland, Roger Temam, Jeffrey Dean, David Grove,
Craig Chambers, Kim B. Bruce, and Elisa Bertino}

\institute{University of Wisconsin-Madison and North Carolina State University
}

\maketitle              

\begin{abstract}
Least Absolute Shrinkage and Selection Operator or the Lasso, introduced by Tibshirani (1996), is a popular estimation procedure in multiple linear regression when underlying design has a sparse structure, because of its property that it sets some regression coefficients exactly equal to 0. In this article, we develop a perturbation bootstrap method and establish its validity in approximating the distribution of the Lasso in heteroscedastic linear regression. We allow the underlying covariates to be either random or non-random. We show that the proposed bootstrap method works irrespective of the nature of the covariates, unlike the resample-based bootstrap of Freedman (1981) which must be tailored based on the nature (random vs non-random) of the covariates. Simulation study also justifies our method in finite samples.  
\end{abstract}

\section{Introduction}
\label{sec:intro}
Consider the multiple linear regression model
\begin{equation}\label{eqn:model1}
y_{i} = \bm{X}'_{i}\bm{\beta} + \epsilon_{i}, \; \;\;\;\;     i = 1,\dots,n,
\end{equation}
where $y_1,\ldots,y_n$ are responses, $\bm{X}_1,\dots,\bm{X}_n$ are design vectors, $\epsilon_1,\ldots,\epsilon_n$ are errors, and $\bm{\beta}=(\beta_1,\ldots, \beta_p)$ is the $p$-dimensional vector of regression parameters. We will consider both the cases when the design vectors are fixed and when they are random, separately. Additionally, we assume that $\epsilon_i$ are independent (but possibly depends on $\bm{X}_i$) when design is fixed and $(\epsilon_i,\bm{X}_i)$ are independent when the design is random. When $p$ is sufficiently large, it is common to approach regression model (\ref{eqn:model1}) with the assumption that the vector $\bm{\beta}$ is sparse, that is the set $\mathcal{A}= \{j:\beta_j\neq 0\}$ has cardinality $p_0 = |\mathcal{A}|$ that is much smaller than $p$, meaning that only a few of the covariates are ``active''. One of the widely used methods under the sparsity assumption is the least absolute shrinkage and selection operator or the Lasso, introduced by Tibshirani (1996). It is defined as the minimizer of $l_1$-penalized least-square criterion function, 
\begin{equation}\label{eqn:lasso}
\bm{\hat{\beta}}_n = \operatorname*{arg\,min}_{\bm{t}}\Bigg[\sum_{i=1}^{n}(y_i - \bm{X}^
{\prime}_i \bm{t})^2
+\lambda_n\sum_{j=1}^{p}|t_{j}|\Bigg],
\end{equation}
where $\lambda_n>0$ is the penalty parameter. Lasso is well suited to the sparse setting because of its property that it sets some regression coefficients exactly equal to 0 and hence it automatically leads to parsimonious model section [cf. Zhao and Yu (2006), Zou (2006) and Wainwright (2009)]. Another important aspect of Lasso is its computational feasibility in high dimensional regression problems [cf. Efron et al. (2004), Friedman et al. (2007), Fu (1998), Osborne et al. (2000)]. 

\par
Asymptotic properties of the Lasso estimator was first investigated by Knight and Fu (2000) [hereafter referred to as KF(00)] for the model (\ref{eqn:model1}) with non-random design and homoscedastic error, when $p$ is assumed fixed. Subsequently, in fixed $p$ setting, Wagener and Dette (2012) and Camponovo (2015) extended results obtained by KF(00) to the cases when errors are heteroscedastic and the design is random. Although it follows that the Lasso estimator is $\sqrt{n}$-consistent for the finite dimensional model (\ref{eqn:model1}), corresponding asymptotic distribution is complicated for constructing confidence regions and perform tests on regression parameter. Hence, a practically reasonable approximation to the distribution of the Lasso estimator is necessary for the purpose of inference. A general alternative approach of approximation besides the asymptotic distribution is the corresponding bootstrap distribution. When the design is non-random and errors are homoscedastic, KF(00) considered the residual bootstrap [cf. Freedman (1981)] in Lasso regression. Chatterjee and Lahiri (2010) investigated the asymptotic properties of usual residual bootstrap in Lasso and found that it fails drastically in approximating the distribution of Lasso when the regression parameter vector is sparse. More precisely, they found that the asymptotic distribution of the residual bootstrapped Lasso estimator is a non-degenerate random measure on $\mathcal{R}^p$ when one or more components of the regression parameter is zero. 

\par
To make residual bootstrap work, Chatterjee and Lahiri (2011) proposed a modification in residual bootstrap method and showed that it is consistent in approximating the distribution of the Lasso estimator when the design is non-random and errors are homoscedastic. Recently, Camponovo (2015) investigated paired bootstrap [cf. Freedman (1981)] in Lasso when the design is random and the errors are heteroscedastic. He pointed out two reasons which are resulting in the failure of the paired bootstrap method and introduced a modification and established its consistency in estimating the distribution of the Lasso estimator.

\par
In this paper, we construct a bootstrap method based on the perturbation approach [cf. Jin et. al (2001), Das and Lahiri (2016)] in both random and non-random design case and when the errors are allowed to be heteroscedastic. See Section 2 for details. We show that the conditional distribution of the corresponding bootstrapped Lasso estimator is consistent in estimating the complicated distribution of the original Lasso estimator, even in sparse set-up. To prove our results, first we show strong consistency of Lasso, extending the result of Chatterjee and Lahiri (2010) for the case of non-random design and homoscedastic errors, and then we employ the techniques of finding asymptotic distribution of the M-estimators with non-differentiable convex objective functions [cf. Pollard (1991), Hjort and Pollard (1993), Geyer (1994, 1996), Kato (2009)], similar to what was employed by KF(00), to find the asymptotic distribution of the original Lasso estimator.  

\par
We conclude this section with a brief literature review. The perturbation bootstrap was introduced by Jin, Ying, and Wei (2001) as a resampling procedure where the objective function has a U-process structure. Work on the perturbation bootstrap in the linear regression setup is limited. Some work has been carried out by Chatterjee and Bose (2005), Minnier et al. (2011), Zhou, Song and Thompson (2012), Das and Lahiri (2016), Das et al. (2017). As a variable selection procedure, Tibshirani (1996) introduced the Lasso. Literature on bootstrap methods in Lasso is very limited. Asymptotic properties of standard residual bootstrap in Lasso has been investigated by Knight and Fu (2000) and Chatterjee and Lahiri (2010). Chatterjee and Lahiri (2010) considered regression model (\ref{eqn:model1}) with non-random covariates and homoscedastic errors and showed that residual bootstrap fails drastically in approximating distribution of Lasso when one or more regression coefficients are zero. Subsequently Chatterjee and Lahiri (2011) developed a thresholding approach which rectifies this problem. Recently Camponovo (2015) developed a modified paired bootstrap for Lasso in regression model (\ref{eqn:model1}) with random covariates and heteroscedastic errors and established its distributional consistency.

\par
The rest of the paper is organized as follows. Our proposed perturbation bootstrap method for Lasso is described in Section \ref{sec:bootstrap}. The naive perturbation bootstrap and its shortcomings are also discussed in this Section. Main results concerning the estimation properties of the perturbation bootstrap are given in Section \ref{sec:mainresults1}. A moderate simulation study is presented in Section \ref{sec:simulation}. The proofs are provided in Section \ref{sec:proofs1}. Section \ref{sec:1.77} states concluding remarks.

\section{Description of the Bootstrap Method}
\label{sec:bootstrap}
This section is divided into two parts. The first sub-section describes the naive way of defining perturbation bootstrapped Lasso estimator and subsequently points out its shortcomings. The second sub-section describes our proposed modification in defining the perturbation bootstrap version of the Lasso estimator. 

\subsection{Naive Perturbation Bootstrap}

The perturbation bootstrap version of the least square estimator is defined as the minimizer of the objective function $\sum_{i=1}^{n}(y_i - \bm{x}'_i \bm{t}^*)^2G^*_i$ [cf. Jin et al. (2001), Das and Lahiri (2016)]. Thus the natural way of defining corresponding Lasso estimator is
\begin{align}
\check{\bm{\beta}}_n^*=\operatorname*{arg\,min}_{\bm{t}^*}\Big[\sum_{i=1}^{n}(y_i - \bm{X}'_i \bm{t}^*)^2G^*_i
+\lambda_n^*\sum_{j=1}^{p}|t_{j}^*|\Big],
\end{align}
similar to how perturbation bootstrap version is defined for adaptive lasso and SCAD estimators in Minnier et al. (2011). Here $\lambda_n^*$ is a penalty parameter having same asymptotic property as $\lambda_n$.
Again $G_1^*,\ldots, G_n^*$ are $n$ independent copies of a non-degenerate non-negative random variable $G^* \in [0,\infty)$ having expectation $\mu_{G^*}$ and variance $\sigma_{G^*}^2$ with $\sigma_{G^*}^2=\mu_{G^*}^2$. Note that the restriction $\sigma_{G^*}^2=\mu_{G^*}^2$, that is on the distribution of $G^*$, is more general than the assumptions $\mu_{G^*}=1$ and $\sigma_{G^*}^2=1$, considered in Minnier et al. (2011). Two immediate choices of the family of distribution of $G^*$ are Exponential$(\lambda)$ and Beta$(\alpha, \beta)$ with $\alpha=(\beta-\alpha)/(\alpha+\beta)$. Other choices can be found easily by investigating the generalized beta family of distributions. 

\par
Now consider a sequence of constants $\{a_n\}_{n\geq 1}$ such that $a_n+(n^{-1/2}\log n)a_n^{-1} \rightarrow 0$ as $n\rightarrow \infty$ and a $\sqrt{n}$-consistent estimator $\breve{\bm{\beta}}_{n}$ of $\bm{\beta}$. Define the modified estimator $\tilde{\bm{\beta}}_n=(\tilde{\beta}_{n,1},\dots,\tilde{\beta}_{n,p})^\prime$ with $\tilde{\beta}_{n,j}=\breve{\beta}_{n,j}\mathbf{1}(|\breve{\beta}_{n,j}|>a_n)$, $\mathbf{1}(\cdot)$ being the indicator function. It follows from the results of KF(00) and Camponovo (2015) that the Lasso estimator $\hat{\bm{\beta}}_{n}$ is $\sqrt{n}$-consistent and hence for the sake of definiteness we take $\breve{\bm{\beta}}_{n}=\hat{\bm{\beta}}_{n}$. We need to use $\sqrt{n}(\check{\bm{\beta}}_n^*-\tilde{\bm{\beta}}_n)$, in place of $\sqrt{n}(\check{\bm{\beta}}_n^*-\hat{\bm{\beta}}_n)$, to approximate the distribution of $\sqrt{n}(\hat{\bm{\beta}}_n-\bm{\beta})$. The reason is that $\tilde{\bm{\beta}}_n$ can capture the signs of the zero components with probability close to $1$ for sufficiently large $n$ whereas the original Lasso estimator $\hat{\bm{\beta}}_n$ can not, as pointed out in Section 2 of Chatterjee and Lahiri (2011). 

\par
Now to point out the shortcomings of the naive perturbation bootstrapped Lasso estimator $\check{\bm{\beta}}_n^*$, define $\check{\bm{u}}_n^*=\sqrt{n}(\check{\bm{\beta}}_n^*-\tilde{\bm{\beta}}_n)$.  Then we have
\begin{align}\label{eqn:npb}
\bm{\check{u}}_n^{*}= \operatorname*{arg\,min}_{\bm{v}^*}\Bigg[\bm{v}^{*\prime}\bm{C}_n^*\bm{v}^{*} -2\bm{v}^{*\prime}\bm{W}_n^{*}+\lambda_n^*\sum_{j=1}^{p}\Big(|\tilde{\beta}_{j,n}+\dfrac{v_{j}^*}{\sqrt{n}}|-|\tilde{\beta}_{j,n}|\Big)\Bigg]
\end{align}
where $\bm{C}_n^*=n^{-1}\sum_{i=1}^{n}\bm{x}_i\bm{x}_i^\prime G_i^*$, $\bm{W}_n^*=n^{-1/2}\sum_{i=1}^{n}\tilde{\epsilon}_i\bm{X}_iG_i^*$ and $\tilde \epsilon_i = y_i - \bm{x}'_i\tilde{\bm{\beta}}_n$, $i\in\{1,\dots,n\}$. Clearly $\bm{W}_n^*$ is a sequence of non-centered random vectors and hence the asymptotic mean of that quantity is not necessarily $\bm{0}$, needed for consistency in estimating the distribution of the Lasso estimator [see in (\ref{eqn:Zn1}) that $\bm{\tilde{W}}_n^*$ is properly centered in $\bm{Z}_n^*(\bm{v}^*)$]. Additionally from computational perspective, it would be plausible if $\bm{C}_n^*$ can be replaced by $\mu_{G^*}\bm{C}_n$, since $\bm{C}_n^*$ needs to be computed at each bootstrap iteration. If we implement the modification described in the next sub-section, then both the theoretical and computational shortcomings of the naive method become resolved and distributional consistency is achieved by perturbation bootstrap.

\subsection{Modified Perturbation Bootstrap}
In order to rectify the naive method, we need to incorporate a perturbed least-squares criterion involving predicted values $\tilde{y}_i = \bm{x}'_i\tilde{\bm{\beta}}_n$, $i\in\{1,\dots,n\}$, along-with the same involving the observed values $y_1,\dots,y_n$ in the objective function. Formally, the modified perturbation bootstrap version of the Lasso estimator is defined as
\begin{align}\label{eqn:boot}
\hat{\bm{\beta}}_n^* = \operatorname*{arg\,min}_{\bm{t}^*}&\Bigg[\sum_{i=1}^{n}(y_i - \bm{X}'_i \bm{t}^*)^2(G^*_i-\mu_{G^*}) \nonumber\\
&+\sum_{i=1}^{n}(\tilde{y}_i-\bm{X}'_i\bm{t}^*)^2(2\mu_{G^*}-G_i^*)+\mu_{G^*}\lambda_n\sum_{j=1}^{p}|t_{j}^*|\Bigg].
\end{align} 

\par
Now we point out an important computational characteristic of our proposed bootstrap method.
For $i=1,\dots,n$, set $z_i=\tilde{y}_i+\tilde{\epsilon}_i\mu_{G^*}^{-1}(G_i^*-\mu_{G^*})$, where $\tilde \epsilon_i = y_i - \tilde{y}_i$. Then the perturbation bootstrap version of the Lasso estimator, $\hat{\bm{\beta}}_n^*$, can be expressed as 
\begin{align}\label{eqn:altboot}
\hat{\bm{\beta}}_n^*=\operatorname*{arg\,min}_{\bm{t}}\Big{[}\sum_{i=1}^{n}\big(z_i-\bm{X}'_i\bm{t}\big)^2+\lambda_n\sum_{j=1}^{p}|t_{j}|\Big{]}
\end{align}
This representation indicates that $\bm{\hat{\beta}_n^*}$ is a Lasso estimator corresponding to the pseudo observations $\{z_1,\dots,z_n\}$ and the design vectors $\bm{x}_1,\dots,\bm{x}_n$. Hence we can employ existing computationally fast algorithms, eg. LARS [cf. Efron et al. (2004)], ``one-at-a-time" coordinate-wise descent [cf. Friedman et al. (2007)], even for finding the perturbation bootstrap version of the Lasso estimator. In the next section, we present the theoretical properties of the proposed perturbation bootstrap method.


\section{Main Results}\label{sec:mainresults1}
\subsection{Notations}
Suppose, $\bm{\beta}=(\beta_1,\dots,\beta_p)^\prime$ denotes the true value of the regression parameter vector. Let, $\mathcal{B}(\mathcal{R}^p)$ denotes the Borel sigma-field defined on $\mathcal{R}^p$ and $\rho(\cdot,\cdot)$ denotes the Prokhorov metric on the collection of all probability measures on $\big(\mathcal{R}^p,\mathcal{B}(\mathcal{R}^p)\big)$.  Set $\mathcal{A}=\{j: \beta_{j}\neq 0\}$ and  $p_0=|\mathcal{A}|$, supposing, without loss of generality, that $\mathcal{A}_n=\{1,\ldots,p_0\}$. $\mathscr{E}$ is the sigma-field generated by $\{\epsilon_i:i\geq 1\}$ when the design is non-random whereas when the design is non-random, $\mathscr{E}$ is the sigma-field generated by $\{(\epsilon_i,\bm{X}_i):i\geq 1\}$. Further assume that $\mathbf{F}_n$ is the distribution of $\bm{T}_n=\sqrt{n}(\hat{\bm{\beta}}_n-\bm{\beta})$. The bootstrap version of $\bm{T}_n$ is $\bm{T}_n^*=\sqrt{n}(\hat{\bm{\beta}}_n^*-\tilde{\bm{\beta}}_n)$ and $\hat{\mathbf{F}}_n$ is the conditional distribution of $\bm{T}_n^*$ given $\mathscr{E}$. Define $\mathbf{F}_{\infty}$ to be the limit distribution of $\bm{T}_n$. Suppose, $\mathbf{P}_*$ and $\mathbf{E}_*$ respectively denote the bootstrap probability and bootstrap expectation conditional on data.

\subsection{Results in the case of non-random designs}
In this subsection, we consider the linear regression model (\ref{eqn:model1}) when the design vectors $\bm{X}_1,\dots,\bm{X}_n$ are fixed. The errors $\epsilon_1,\dots,\epsilon_n$ are independent; but may not be identically distributed. Thus $\epsilon_i$ may be assumed to depend on $\bm{X}_i$, $i\in\{1,\dots,n\}$. In this setup, we show that $\hat{\mathbf{F}}_n$ is a valid approximation of $\mathbf{F}_n$, as formally stated in the next theorem.

\begin{theo}\label{eqn:theo1}
Suppose the following regularity conditions hold:
\begin{enumerate}[label=(A.\arabic*)]
\item $n^{-1}\sum_{i=1}^{n}\bm{X}_i\bm{X}_i^\prime\rightarrow \bm{C}$ as $n\rightarrow \infty$, for some positive definite matrix $\bm{C}$.
\item $n^{-1}\sum_{i=1}^{n}||\bm{X}_i||^{4+\delta}=O(1)$ for some $\delta>0$.
\item $\mathbf{E}\epsilon_i=0$ for all $i\in\{1,\dots,n\}$ and $n^{-1}\sum_{i=1}^{n}\mathbf{E}|\epsilon_i|^{4+\delta}=O(1)$.
\item $n^{-1}\sum_{i=1}^{n}\bm{X}_i\bm{X}_i^\prime \mathbf{E}\epsilon_i^2\rightarrow \bm{\Sigma}$ as $n\rightarrow \infty$, for some positive definite matrix $\bm{\Sigma}$. 
\item $\lambda_n/\sqrt{n}\rightarrow \lambda_0 \in [0,\infty)$.
\item $G_{i}^{*}$ and $\epsilon_i$ are independent for all $i\in\{1,\dots,n\}$ and $\mathbf{E}G_1^{*3}<\infty$.
\end{enumerate}
Then it follows that $\rho(\hat{\mathbf{F}}_n,\mathbf{F}_n)\rightarrow 0$ as $n\rightarrow \infty$, with probability $1$.
\end{theo}

\subsubsection{Incompetency of Residual Bootstrap when Errors are Heteroscedastic}
First, let us briefly describe the residual bootstrap method in Lasso, developed in Chatterjee and Lahiri (2011). The modified residuals
$\{\tilde{\epsilon}_1,\dots,\tilde{\epsilon}_n\}$ by $\tilde{\epsilon}_i = y_i - \bm{X}_i^\prime\tilde{\beta}_n$. Suppose $\bar{\epsilon}_n$ is the mean of the residuals. Then select a random sample $\{r_1^*,\dots,r_n^*\}$ from $\{(\tilde{\epsilon}_1-\bar{\epsilon}_n),\dots,(\tilde{\epsilon}_n-\bar{\epsilon}_n)\}$ and define
\begin{align*}
y_i^* = \bm{X}_i^\prime\tilde{\beta}_n + r_i^*\;\;\;\;\;, i=1,\dots,n
\end{align*}
Then the residual bootstrapped Lasso estimator is defined as
\begin{align}\label{eq:relasso}
\breve{\bm{\beta}}_n^*=\operatorname*{arg\,min}_{\bm{t}^*}\Big[\sum_{i=1}^{n}(y_i^* - \bm{X}'_i \bm{t}^*)^2
+\lambda_n\sum_{j=1}^{p}|t_{j}^*|\Big].
\end{align}
To prove incompetency of residual bootstrap in Lasso, our argument is similar to as in Liu (1988) for least square estimator (LSE). When $p=1$, Liu (1988) showed that the conditional variance of residual bootstrapped LSE is not a consistent estimator of the variance of LSE. Wagener and Dette (2011) showed that under the conditions (A.1), (A.4), (A.5), $\mathbf{E}\epsilon_i=0$ for all $i\in\{1,\dots,n\}$ and $n^{-1}\max_i||\bm{X}_i||^2\sigma_i^2\rightarrow 0$ [Note that this condition follows from (A.2) and (A.3)]
\begin{align*}
\sqrt{n}\big(\hat{\bm{\beta}}_n-\bm{\beta}\big)\xrightarrow{d} \operatorname*{arg\,min}_{\bm{v}} Z(\bm{v})
\end{align*}
where 
\begin{align*}
Z(\bm{v})=\Bigg[\bm{v}^\prime\bm{C}\bm{v} -2\bm{v}^{\prime}\bm{W}+\lambda_0\Big(\sum_{j=1}^{p_0}sgn({{\beta}_{j}})v_j+\sum_{j=p_0+1}^{p}|v_j|\Big)\Bigg]
\end{align*}
where $\bm{W}$ follows $N(\bm{0},\bm{\Sigma})$, when $\mathcal{A}=\{1,\dots,p_0\}$. The following proposition implies the incompetency of residual bootstrap in case of Lasso when errors are heteroscedastic.

\begin{prop}\label{prop:resi}
Suppose, conditions (A.1)-(A.3) and (A.5) hold. Also assume that $n^{-1}\sum_{i=1}^{n}$ $\mathbf{E}\epsilon_i^2\rightarrow s^2$ $(>0)$. Then the following is true. 
\begin{align*}
&\sqrt{n}\big(\breve{\bm{\beta}}^*_n-\tilde{\bm{\beta}}\big)\xrightarrow{d} \operatorname*{arg\,min}_{\bm{v}} \tilde{Z}(\bm{v})\\
\text{with}\;\;& \tilde{Z}(\bm{v})=\Bigg[\bm{v}^\prime\bm{C}\bm{v} -2\bm{v}^{\prime}\tilde{\bm{W}}+\lambda_0\Big(\sum_{j=1}^{p_0}sgn({{\beta}_{j}})v_j+\sum_{j=p_0+1}^{p}|v_j|\Big)\Bigg] 
\end{align*}
where $\tilde{\bm{W}}$ follows $N(\bm{0},s^2\bm{C})$.
\end{prop}

\begin{remark}
Note that $\bm{\Sigma}=s^2\bm{C}$ when the errors are homoscedastic and hence $\tilde{Z}(\bm{v})$ becomes $Z(\bm{v})$. Therefore residual bootstrap works in homoscedastic setting, as found in Chatterjee and Lahiri (2011). But when the errors become heteroscedastic, then residual bootstrap is no longer valid in approximating the distribution of Lasso as long as $\bm{\Sigma}\neq s^2\bm{C}$. This is indeed the case in most of the situations whenever the variance of the errors depend on the covariates. On the other hand, perturbation bootstrap can consistently approximate even when the regression errors are no longer homoscedastic. This is an advantage of perturbation bootstrap over the residual bootstrap.
\end{remark}

\subsection{The Result in the case of random designs}
In this subsection, we establish that $\hat{\mathbf{F}}_n$ approximates $\mathbf{F}_n$ consistently when the design vectors are random and errors $\epsilon_1,\dots,\epsilon_n$ are independent, but not necessarily identically distributed. The result is stated in the theorem \ref{eqn:theo2} below.

\begin{theo}\label{eqn:theo2}
Suppose the following regularity conditions hold:
\begin{enumerate}[label=(B.\arabic*)]
\item $\mathbf{E}\Big[n^{-1}\sum_{i=1}^{n}\bm{X}_i\bm{X}_i^\prime\Big]\rightarrow \bm{C}$ as $n\rightarrow \infty$, for some positive definite matrix $\bm{C}$.
\item $n^{-1}\sum_{i=1}^{n}\mathbf{E}||\bm{X}_i||^{4+\delta}=O(1)$ for some $\delta>0$.
\item $\mathbf{E}(\epsilon_i|\bm{X}_i)=0$ for all $i\in\{1,\dots,n\}$ and $n^{-1}\sum_{i=1}^{n}\mathbf{E}|\epsilon_i|^{4+\delta}=O(1)$.   
\item $\mathbf{E}\Big{(}n^{-1}\sum_{i=1}^{n}\bm{X}_i\bm{X}_i^\prime\epsilon_i^2\Big{)}\rightarrow \bm{\Sigma}$ as $n\rightarrow \infty$, for some positive definite matrix $\bm{\Sigma}$. 
\item $\lambda_n/\sqrt{n}\rightarrow \lambda_0 \in [0,\infty)$.
\item $G_{i}^{*}$ and $(\epsilon_i,\bm{X}_i)$ are independent for all $i\in\{1,\dots,n\}$ and $\mathbf{E}G_1^{*3}<\infty$.
\end{enumerate}
Then it follows that $\rho(\hat{\mathbf{F}}_n,\mathbf{F}_n)\rightarrow 0$ as $n\rightarrow \infty$, with probability $1$.
\end{theo}

\begin{remark}
Theorem \ref{eqn:theo1} and \ref{eqn:theo2} establishes distributional consistency of our proposed bootstrap method in approximating the distribution of Lasso. Formally, it follows that on a set of probability $1$, $\hat{\mathbf{F}}_n\rightarrow \mathbf{F}_{\infty}$ in distribution and hence $\mathbf{F}(B)-\hat{\mathbf{F}}(B)\rightarrow 0$ as $n\rightarrow \infty$ for all $B\in \mathcal{R}^p$ with $\bm{F}_{\infty}(\partial B)=0$. Since the form of the asymptotic distribution of $\hat{\bm{\beta}}$ is quite complicated [cf. Chatterjee and Lahiri (2011), Camponovo (2015)], validity of perturbation bootstrap approximation, as stated in the theorems, is essential if one wants to infer about $\bm{\beta}$ based on the Lasso estimator $\hat{\bm{\beta}}_n$.
\end{remark}

\begin{remark}
One needs to implement paired or residual bootstrap depending on whether the covariates are random or non-random and also there is significant difference in implementation between these two procedures. On the other hand Theorem \ref{eqn:theo1} and \ref{eqn:theo2} are implying that one can implement perturbation bootstrap without thinking about the nature of the covariates. This means that the perturbation bootstrap is more general than the resample-based bootstrap (residual and paired) of Freedman (1981) in case of Lasso. 
\end{remark}


\section{Simulation results}
\label{sec:simulation}

We study through simulation the coverage of one-sided and two-sided $90\%$ confidence intervals for individual regression coefficients separately for both the situations when design is non-random and random, constructed using Theorem \ref{eqn:theo1} and \ref{eqn:theo2}. We compare empirical coverages of our perturbation bootstrap (PB) method with that of residual (RB) and paired bootstrap (PaB). We have also compared the empirical coverage of confidence region obtained by perturbation bootstrap with other two methods. 

\par

\begin{table}[ht]
\centering
\addtolength{\tabcolsep}{4pt} 
\begin{tabular}{r|cc|cc}
\multicolumn{5}{l}{Coverage and \textit{(avg.~width)} of two-sided 90\% CIs}\\ \hline\hline 
\multirow{2}{*}{$\beta_j$} &\multicolumn{2}{c|}{Case I}&\multicolumn{2}{c}{Case II}\\
\cline{2-5}
 & $PB$ & $RB$ & $PB$ & $RB$  \\ 
 \hline \hline
1 & 0.598 & 0.518 & 0.789 & 0.712\\
   & \textit{(1.654)} & \textit{(1.283)} & \textit{(1.061)} & \textit{(0.850)}  \\ 
  1.25 & 0.710 & 0.473 & 0.784 & 0.602  \\ 
   & \textit{(1.981)} & \textit{(1.215)} & \textit{(1.187)} & \textit{(0.786)}   \\ 
  1.50 & 0.714 & 0.647 & 0.834 & 0.771  \\ 
   & \textit{(2.018)} & \textit{(1.747)} & \textit{(1.228)} & \textit{(1.052)}   \\ 
  1.75 & 0.763 & 0.733 & 0.856 & 0.829  \\
   & \textit{(1.743)} & \textit{(1.650)} & \textit{(1.006)} & \textit{(0.933)} \\  
   2.00 & 0.760 & 0.764 & 0.893 & 0.900  \\
   & \textit{(1.716)} & \textit{(1.785)} & \textit{(0.986)} & \textit{(1.024)}  \\ 
   2.25 & 0.805 & 0.781 & 0.885 & 0.858  \\
   & \textit{(1.829)} & \textit{(1.833)} & \textit{(1.037)} & \textit{(1.034)}  \\ 
   0 & 0.800 & 0.783 & 0.863 & 0.814  \\
   & \textit{(1.066)} & \textit{(0.927)} & \textit{(0.629)} & \textit{(0.547)}  \\ 
    0 & 0.845 & 0.782 & 0.873 & 0.805  \\
   & \textit{(1.087)} & \textit{(0.891)} & \textit{(0.627)} & \textit{(0.525)}  \\
    0 & 0.843 & 0.755 & 0.855 & 0.743\\
   & \textit{(1.157)} & \textit{(0.877)} & \textit{(0.691)} & \textit{(0.526)} \\
    0 & 0.861 & 0.685 & 0.852 & 0.706  \\
   & \textit{(1.381)} & \textit{(0.901)} & \textit{(0.802)} & \textit{(0.525)} \\\hline \hline
 \multicolumn{5}{l}{ Coverage of one-sided 90\% CIs}\\\hline\hline
1 & 0.629 & 0.626 & 0.824  & 0.796\\ 
  1.25 & 0.701 & 0.671 & 0.822 & 0.751\\ 
  1.50 & 0.769 & 0.763 & 0.863 & 0.827 \\ 
  1.75 & 0.801 & 0.807 & 0.876  & 0.859\\   
  2.00 & 0.822 & 0.826 & 0.893 & 0.900 \\   
  2.25 & 0.837 & 0.847 & 0.864 &  0.870\\   
  0 & 0.873 & 0.852 & 0.878  & 0.856\\   
  0 & 0.733 & 0.697 & 0.847 &  0.794\\   
  0 & 0.832 & 0.753 & 0.840 &  0.798\\   
  0 & 0.865 & 0.731 & 0.854  & 0.771\\      
\end{tabular}
\caption{Empirical coverage of 90\% confidence intervals for regression coefficients over $1000$ simulations under $(n,p,p_0)=(100,10,6)$ using crossvalidation choice of $\lambda_n$. 
One-sided intervals are right sided.} 
\label{tab:p10n100cvlambdanr}
\end{table}

\clearpage

\par
Under the settings 
\[
(n,p,p_0) \in \left\{ (100,10,6), (500,10,6), (1000,10,6) \right\},
\]
we generate $n$ independent copies $(X_1,Y_1),\dots,(X_n,Y_n)$ of $(X,Y) \in \mathcal{R}^p \times \mathcal{R}$ from the model $Y = X^\prime\bbeta + \epsilon$,
where $\epsilon=(\epsilon_1,\dots,\epsilon_n)$. We have considered two cases in both non-random and random design setup. Case I is when $\epsilon_i/s_i$ is generated from $(\chi^2_2-2)$ distribution and case II

\begin{table}[ht]
\centering
\addtolength{\tabcolsep}{4pt} 
\begin{tabular}{r|cc|cc}
\multicolumn{5}{l}{Coverage and \textit{(avg.~width)} of two-sided 90\% CIs}\\ \hline\hline 
\multirow{2}{*}{$\beta_j$} &\multicolumn{2}{c|}{Case I}&\multicolumn{2}{c}{Case II}\\
\cline{2-5}
 & $PB$ & $RB$ & $PB$ & $RB$  \\ 
 \hline \hline
1 & 0.881 & 0.840 & 0.890 & 0.801\\
   & \textit{(3.124)} & \textit{(2.565)} & \textit{(1.847)} & \textit{(1.502)}  \\ 
  1.25 & 0.884 & 0.828 & 0.893 & 0.818  \\ 
   & \textit{(3.562)} & \textit{(2.616)} & \textit{(1.957)} & \textit{(1.484)}   \\ 
  1.50 & 0.914 & 0.841 & 0.899 & 0.848  \\ 
   & \textit{(3.665)} & \textit{(3.133)} & \textit{(2.004)} & \textit{(1.741)}   \\ 
  1.75 & 0.895 & 0.858 & 0.904 & 0.852  \\
   & \textit{(3.331)} & \textit{(3.054)} & \textit{(1.770)} & \textit{(1.620)} \\  
   2.00 & 0.907 & 0.828 & 0.914 & 0.861  \\
   & \textit{(3.300)} & \textit{(3.172)} & \textit{(1.795)} & \textit{(1.719)}  \\ 
   2.25 & 0.893 & 0.760 & 0.895 & 0.845 \\
   & \textit{(3.473)} & \textit{(3.182)} & \textit{(1.781)} & \textit{(1.702)}  \\ 
   0 & 0.908 & 0.825 & 0.901 & 0.847  \\
   & \textit{(2.122)} & \textit{(1.798)} & \textit{(1.149)} & \textit{(0.969)}  \\ 
    0 & 0.892 & 0.815 & 0.893 & 0.806  \\
   & \textit{(2.238)} & \textit{(1.742)} & \textit{(1.189)} & \textit{(0.942)}  \\
    0 & 0.909 & 0.782 & 0.916 & 0.845\\
   & \textit{(2.386)} & \textit{(1.753)} & \textit{(1.246)} & \textit{(0.975)} \\
    0 & 0.920 & 0.747 & 0.886 & 0.821  \\
   & \textit{(2.535)} & \textit{(1.764)} & \textit{(1.391)} & \textit{(0.955)} \\\hline \hline
 \multicolumn{5}{l}{ Coverage of one-sided 90\% CIs}\\\hline\hline
1 & 0.875 & 0.848 & 0.882  & 0.838\\ 
  1.25 & 0.889 & 0.854 & 0.893 & 0.848\\ 
  1.50 & 0.895 & 0.859 & 0.891 & 0.865 \\ 
  1.75 & 0.891 & 0.874 & 0.892  & 0.862\\   
  2.00 & 0.904 & 0.862 & 0.910 & 0.876 \\   
  2.25 & 0.908 & 0.825 & 0.900 &  0.863\\   
  0 & 0.881 & 0.848 & 0.915  & 0.877\\   
  0 & 0.901 & 0.836 & 0.888 &  0.831\\   
  0 & 0.882 & 0.839 & 0.912 &  0.849\\   
  0 & 0.895 & 0.803 & 0.882  & 0.850\\      
\end{tabular}
\caption{Empirical coverage of 90\% confidence intervals for regression coefficients over $1000$ simulations under $(n,p,p_0)=(1000,10,6)$ using crossvalidation choice of $\lambda_n$. 
One-sided intervals are right sided.} 
\label{tab:p10n1000cvlambdanr}
\end{table}
\hspace*{-5mm}is when $\epsilon_i/s_i$ is generated from standard normal distribution with $s_i^2=p^{-1}\sum_{j=1}^{p}|X_{ij}|^5$ where $X_{ij}$ is the $(i,j)$th element of design matrix $\bm{X}$. The design vectors $X_i = (X_{i1},\dots,X_{ip})^\prime$, $i\in\{1,\dots,n\}$, are independent and identically distributed mean-zero multivariate normal random vector such that
\[
\text{Cov}(X_{1j},X_{1k}) = \mathbf{1}(j=k) + 0.3^{|j-k|}\mathbf{1}(j\leq p_0)\mathbf{1}(k \leq p_0)\mathbf{1}(j\neq k)
\] 
for $1 \leq j,k \leq p$, and $\bbeta = (\beta_1,\dots,\beta_p)^\prime$ with $\beta_j$ defined as $\beta_j = (3/4) + (1/4)j$ for $j=1,\dots,p_0$ and $\beta_j = 0$ for $j=p_0+1,\dots,p$.

\begin{table}[ht]
\centering
\addtolength{\tabcolsep}{4pt} 
\begin{tabular}{r|cc|cc}
\multicolumn{5}{l}{Coverage and \textit{(avg.~width)} of two-sided 90\% CIs}\\ \hline\hline 
\multirow{2}{*}{$\beta_j$} &\multicolumn{2}{c|}{Case I}&\multicolumn{2}{c}{Case II}\\
\cline{2-5}
 & $PB$ & $PaB$ & $PB$ & $PaB$  \\ 
 \hline \hline
   1 & 0.642 & 0.683 & 0.827 & 0.834                 \\
   & \textit{(1.408)} & \textit{(1.580)} & \textit{(0.883)} & \textit{(0.944)}  \\ 
  1.25 & 0.704 & 0.723 & 0.874 & 0.875  \\ 
   & \textit{(1.562)} & \textit{(1.723)} & \textit{(0.940)} & \textit{(0.995)}   \\ 
  1.50 & 0.747 & 0.774 & 0.873 & 0.873  \\ 
   & \textit{(1.684)} & \textit{(1.823)} & \textit{(0.954)} & \textit{( 1.004)}   \\ 
  1.75 & 0.781 & 0.800 & 0.852 & 0.856 \\
   & \textit{(1.709)} & \textit{(1.845)} & \textit{(0.952)} & \textit{(1.003)} \\  
   2.00 & 0.815 & 0.840 & 0.852 &  0.851 \\
   & \textit{(1.749)} & \textit{(1.888)} & \textit{(0.956)} & \textit{(1.005)}  \\ 
   2.25 & 0.831 & 0.856 & 0.843 & 0.845\\
   & \textit{(1.696)} & \textit{(1.826)} & \textit{(0.932)} & \textit{(0.979)}  \\ 
   0 & 0.861 & 0.852 & 0.866 &  0.810 \\
   & \textit{(1.137)} & \textit{(1.219)} & \textit{(0.641)} & \textit{(0.675)}  \\ 
    0 & 0.854 & 0.820 & 0.865 & 0.790  \\
   & \textit{(1.157)} & \textit{(1.251)} & \textit{( 0.645)} & \textit{(0.678)}  \\
    0 & 0.871 & 0.849 & 0.869 & 0.815\\
   & \textit{(1.107)} & \textit{(1.196)} & \textit{(0.630)} & \textit{(0.663)} \\
    0 & 0.876 & 0.847 & 0.867 & 0.787 \\
   & \textit{(1.161)} & \textit{(1.241)} & \textit{(0.642)} & \textit{(0.672)} \\\hline \hline
 \multicolumn{5}{l}{ Coverage of one-sided 90\% CIs}\\\hline\hline
   1 & 0.679 & 0.755 & 0.832 & 0.854   \\ 
  1.25 & 0.749 & 0.776 & 0.870 &  0.886   \\ 
  1.50 & 0.806 &  0.831 & 0.869 & 0.890 \\ 
  1.75 & 0.828 & 0.852 & 0.867  & 0.875\\   
  2.00 & 0.824 & 0.861 & 0.867 &  0.882\\   
  2.25 & 0.828 & 0.860 & 0.849 & 0.869\\   
  0 &  0.828 &  0.817 & 0.878 & 0.798 \\   
  0 & 0.827 & 0.830 & 0.860 & 0.790\\   
  0 & 0.848 & 0.843 & 0.873 & 0.800\\   
  0 & 0.819 & 0.832 & 0.870 & 0.785\\      
\end{tabular}
\caption{Empirical coverage of 90\% confidence intervals for regression coefficients over $1000$ simulations under $(n,p,p_0)=(100,10,6)$ using crossvalidation choice of $\lambda_n$. 
One-sided intervals are right sided.} 
\label{tab:p10n100cvlambdar}
\end{table}

\par
We compute the empirical coverage over $1000$ simulated data sets of one- and two-sided confidence intervals for each regression coefficient under crossvalidation-selected values of $\lambda_n$. To construct the bootstrap intervals for each of the 1000 simulated data sets, we generate 1200 independent samples from $Exp(1)$. The values of $\lambda_n$ are thereafter held fixed for all bootstrap computations on the same dataset.

\clearpage

\par
Tables \ref{tab:p10n100cvlambdanr} and \ref{tab:p10n1000cvlambdanr} are displaying the empirical coverage probabilities of 90\% CIs for each of the regression coefficients under the settings $(n,p,p_0) \in \{(100,10,6),(1000,10,6)\}$, when the design is non-random. For the random design case, Tables \ref{tab:p10n100cvlambdar} and \ref{tab:p10n1000cvlambdar} are displaying the empirical coverages when $(n,p,p_0) \in \{(100,10,6),(1000,10,6)\}$. Tables \ref{tab:p10n100cvlambdanr} and \ref{tab:p10n1000cvlambdanr} are comparing our perturbation bootstrap method with residual bootstrap and Tables \ref{tab:p10n100cvlambdar} and \ref{tab:p10n1000cvlambdar} are comparing our perturbation bootstrap method with paired bootstrap.

\begin{table}[ht]
\centering
\addtolength{\tabcolsep}{4pt} 
\begin{tabular}{r|cc|cc}
\multicolumn{5}{l}{Coverage and \textit{(avg.~width)} of two-sided 90\% CIs}\\ \hline\hline 
\multirow{2}{*}{$\beta_j$} &\multicolumn{2}{c|}{Case I}&\multicolumn{2}{c}{Case II}\\
\cline{2-5}
 & $PB$ & $PaB$ & $PB$ & $PaB$  \\ 
 \hline \hline
1 & 0.910 & 0.907 & 0.900 & 0.890        \\
   & \textit{(2.911)} & \textit{(3.106)} & \textit{(1.668)} & \textit{(1.733)}  \\ 
  1.25 & 0.895 & 0.893 & 0.910 & 0.903  \\ 
   & \textit{(3.131)} & \textit{(3.307)} & \textit{(1.746)} & \textit{(1.804)}   \\ 
  1.50 & 0.903 & 0.903 & 0.892 & 0.889  \\ 
   & \textit{(3.251)} & \textit{(3.404)} & \textit{(1.762)} & \textit{(1.815)}   \\ 
  1.75 & 0.909 & 0.906 & 0.899 & 0.898 \\
   & \textit{(3.289)} & \textit{(3.437)} & \textit{(1.758)} & \textit{(1.812)} \\  
   2.00 & 0.891 & 0.894 & 0.904 &  0.903 \\
   & \textit{(3.333)} & \textit{(3.487)} & \textit{(1.762)} & \textit{(1.814)}  \\ 
   2.25 & 0.903 & 0.901 & 0.907 & 0.901\\
   & \textit{(3.227)} & \textit{(3.369)} & \textit{(1.716)} & \textit{(1.766)}  \\ 
   0 & 0.910 & 0.906 & 0.900 &  0.784 \\
   & \textit{(2.207)} & \textit{(2.322)} & \textit{(1.189)} & \textit{(1.233)}  \\ 
    0 & 0.908 & 0.902 & 0.917 & 0.776  \\
   & \textit{(2.235)} & \textit{(2.359)} & \textit{( 1.195)} & \textit{(1.238)}  \\
    0 & 0.897 & 0.892 & 0.892 & 0.755\\
   & \textit{(2.172)} & \textit{(2.295)} & \textit{(1.181)} & \textit{(1.223)} \\
    0 & 0.907 & 0.898 & 0.890 &   0.752 \\
   & \textit{(2.238)} & \textit{(2.351)} & \textit{(1.193)} & \textit{(1.233)} \\\hline \hline
 \multicolumn{5}{l}{ Coverage of one-sided 90\% CIs}\\\hline\hline
   1 & 0.884 & 0.886 & 0.904  & 0.901     \\ 
  1.25 & 0.898 & 0.902 & 0.891 &  0.893   \\ 
  1.50 & 0.896 &  0.899 & 0.894 & 0.900 \\ 
  1.75 & 0.901 & 0.913 & 0.897  & 0.897\\   
  2.00 & 0.895 & 0.898 & 0.906 &  0.911\\   
  2.25 & 0.897 & 0.900 & 0.900 & 0.899\\   
  0 &  0.897 &  0.886 & 0.907 & 0.768 \\   
  0 & 0.904 & 0.891 & 0.920 & 0.790\\   
  0 & 0.892 & 0.884 & 0.890 & 0.763\\   
  0 & 0.899 & 0.889 & 0.911 & 0.765\\      
\end{tabular}
\caption{Empirical coverage of 90\% confidence intervals for regression coefficients over $1000$ simulations under $(n,p,p_0)=(1000,10,6)$ using crossvalidation choice of $\lambda_n$. 
One-sided intervals are right sided.} 
\label{tab:p10n1000cvlambdar}
\end{table}

\begin{figure}[ht]
\begin{center}
\includegraphics[width=0.70\textwidth]{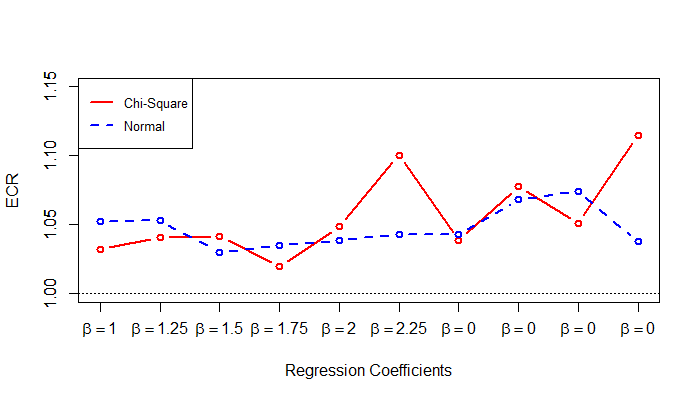}
\end{center}
\caption{Ratio of Empirical Coverage of 90\% CIs constructed by Perturbation and Residual Bootstrap for Regression Coefficients}
\label{fig:riboflavin1}
\end{figure}


\begin{figure}[ht]
\begin{center}
\includegraphics[width=0.70\textwidth]{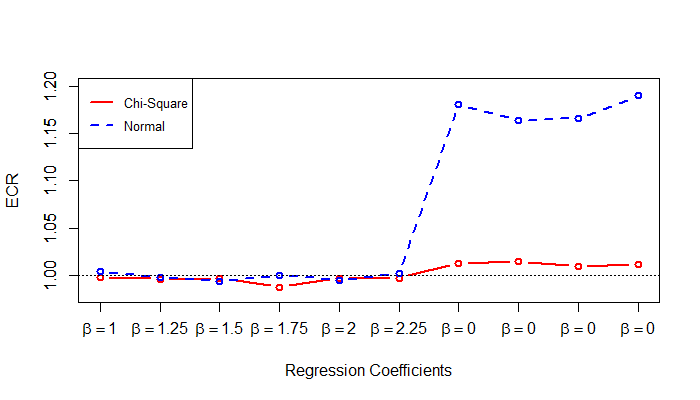}
\end{center}
\caption{Ratio of Empirical Coverage of 90\% CIs constructed by Perturbation and Paired Bootstrap for Regression Coefficients}
\label{fig:riboflavin2}
\end{figure}

\par

In the non-random design case, for which Table \ref{tab:p10n100cvlambdanr} and \ref{tab:p10n1000cvlambdanr} display the empirical coverages of 90\% CIs, the perturbation bootstrap intervals based on $\T_n^*$ perform much better than residual bootstrap intervals for both zero and non-zero regression coefficients. When $n=1000$, perturbation bootstrap two-sided and one-sided intervals achieve closest-to-nominal coverage in both the cases. The empirical coverages of 90\% residual bootstrap CIs, specially the two-sided ones, continue to perform poorly even when the sample size increases from 100 to 1000. These finite sample results are justifying the incompetency of the residual bootstrap when errors are heteroscedastic, as stated in proposition \ref{prop:resi}. When $n=100$ and errors are heteroscedastic normal, the one- and two-sided perturbation bootstrap intervals achieve sub-nominal coverage for the smaller non-zero regression coefficients given that those coefficients were occasionally estimated to be zero, but achieves close-to-nominal coverage for the larger non-zero regression coefficients. For the zero coefficients, the difference in empirical coverages are higher than the non-zero ones in both case I and II, showing that the thresholding, performed on original Lasso estimator to get proper centering in bootstrap setup, has significant influence in achieving nominal coverage in case of perturbation bootstrap than the same in case of residual bootstrap, when the errors are heteroscedastic.

Table \ref{tab:p10n100cvlambdar} and \ref{tab:p10n1000cvlambdar} display the empirical coverages of 90\% CIs in the random design case. In case I, that is when the errors are heteroscedastic centered $\chi^2_2$, the perturbation bootstrap intervals based on $\T_n^*$ perform similarly to paired bootstrap intervals for both zero and non-zero regression coefficients. In case II, although the performance of perturbation and paired bootstrap are equivalent for the non-zero regression coefficients, the empirical coverages of the paired bootstrap are worse than that of the perturbation bootstrap intervals even when $n=1000$. Perturbation and paired bootstrap are performing almost similarly in random design since both can consistently approximate the distribution of Lasso even when the errors are heteroscedastic.

To compare our perturbation bootstrap method with the existing bootstrap methods, we define the quantity ``Empirical Coverage Ratio'' ($ECR$) as 
\begin{align*}
ECR = \dfrac{\text{Empirical Coverage of a Perturbation Bootstrap CI}}{\text{Empirical Coverage of corresponding CI constructed using other bootstrap method}}
\end{align*}
Figures \ref{fig:riboflavin1} and \ref{fig:riboflavin2} are respectively displaying $ECR$ corresponding to the one-sided 90\% CIs constructed using the residual and the paired bootstrap when $n=1000$. Note that $ECR$ values displayed in the Figure \ref{fig:riboflavin1} are greater than $1$ for each of the regression coefficients in both the cases and hence the perturbation bootstrap method is performing better than its residual counterpart in terms of coverages of 90\% one-sided CIs when $n=1000$. Conclusions will be similar for other two sample sizes also, for $n=100$ see Table \ref{tab:p10n100cvlambdanr}. Figure \ref{fig:riboflavin2} is indicating that the perturbation and the paired bootstrap are equivalent in our finite sample setting for non zero regression coefficients in both the cases. But for zero regression coefficients, $ECR$ values are greater than $1$ when errors are heteroscedastic mean zero Normal and and greater than $1.15$ when errors are heteroscedastic centered $\chi^2_2$. Therefore perturbation bootstrap is performing better than paired bootstrap uniformly for all the zero coefficients when $n=1000$. Same is true for $n=100$ also, as can be seen in Table \ref{tab:p10n100cvlambdar}. 

\begin{table}[ht]
\centering
\addtolength{\tabcolsep}{4pt} 
\begin{tabular}{r|cc|cc}
\multicolumn{5}{l}{Empirical Coverage of 90\% Coverage Region}\\ \hline\hline 
\multirow{2}{*}{$n$} &\multicolumn{2}{c|}{Case I}&\multicolumn{2}{c}{Case II}\\
\cline{2-5}
 & $PB$ & $RB$ & $PB$ & $RB$  \\ 
 \hline \hline
   100 & 0.893 & 0.683 & 0.916 & 0.663       \\ 
  500 & 0.932 & 0.773 & 0.942 & 0.806   \\ 
  1000 & 0.948 &  0.754 & 0.932 & 0.795 \\\hline \hline
\multirow{2}{*}{$n$} &\multicolumn{2}{c|}{Case I}&\multicolumn{2}{c}{Case II}\\
\cline{2-5}
 & $PB$ & $PaB$ & $PB$ & $PaB$  \\ 
 \hline \hline
   100 & 0.894 & 0.875 & 0.924 & 0.880         \\ 
  500 & 0.946 & 0.896 & 0.928 & 0.881   \\ 
  1000 & 0.943 &  0.895 & 0.938 & 0.879 \\    
\end{tabular}
\caption{Empirical coverage of 90\% confidence region for regression parameter vector over $1000$ simulations under $(n,p,p_0)=(1000,10,6)$ using crossvalidation choice of $\lambda_n$.} 
\label{tab:p10CRcvlambdarnr}
\end{table}

Table \ref{tab:p10CRcvlambdarnr} is displaying the empirical coverages of 90\% two-sided confidence regions, constructed using perturbation, residual and paired bootstrap, under $(n,p,p_0) \in \{ (100,10,6),$ $ (500,10,6), (1000,10,6) \}$ in both the cases. Whereas perturbation bootstrap empirical coverages are little higher than 0.90 for $n=100$ and 500, residual bootstrap is performing poorly for all the three sample sizes and in both choices of error distribution. On the other hand paired bootstrap is performing nominally.

We see that the perturbation bootstrap is able to produce reliable confidence intervals for regression coefficients and that it is able to do so under cross validation choice of penalty parameter.


\par

\section{Proofs}\label{sec:proofs1}

Suppose, $K$, $K_1$, $K_2$ denote generic constants which do not depend on $n$. Let, $(C.1)-(C.6)$ refers to $(A.1)-(A.6)$ when the design matrix is non-random and to $(B.1)-(B.6)$ when the design matrix is random. Recall that $\mathbf{P}_*$ denotes the conditional probability given $\mathscr{E}$ and $\mathbf{E}_*(\cdot)=\mathbf{E}(\cdot|\mathscr{E})$. Define $\bm{C}_n=n^{-1}\sum_{i=1}^{n}\bm{X}_i\bm{X}'_i$ and $\bm{\tilde{W}}_n^*=n^{-1/2}\sum_{i=1}^{n}\tilde{\epsilon}_i\bm{X}_i(G_i^*-\mu_{G^*})$. For a random vector $\bm{Z}$ and a sigma-field $\mathcal{C}$, suppose $\mathcal{L}(\bm{Z})$ denotes the distribution of $\bm{Z}$ and $\mathcal{L}(\bm{Z}|\mathcal{C})$ denotes the conditional distribution of $\bm{Z}$ given $\mathcal{C}$. For simplicity set $\mathcal{L}(\bm{Z}|\sigma(\bm{W}))=\mathcal{L}(\bm{Z}|\bm{W})$. Also suppose that $||\cdot||$ is the euclidean norm. We will write $w.p.$ to denote ``with probability''. $``\xrightarrow{d}$ $"$ denotes the convergence in distribution.

Now we state some lemmas before proving theorem \ref{eqn:theo1} and \ref{eqn:theo2}.

\begin{lem}\label{lem:indcon}
Let $\{Z_n\}$ be a sequence of independent random variables such that $n^{-\alpha_n}\sum_{i=1}^{n}$ $\mathbf{E}|X_n|^{\alpha_n}$ $=O(1)$ as $n\rightarrow \infty$ for some $\alpha_n\in [1,2]$. Then as $n\rightarrow \infty$,
\begin{align*}
n^{-1}\sum_{i=1}^{n}(Z_j-\mathbf{E}Z_j)=o(1),\;\;\text{w.p.}\; 1 .
\end{align*}
\end{lem}

The above Lemma is stated as Theorem 8.4.6 in Atreya and Lahiri (2006).

\begin{lem}\label{lem:con}
Suppose conditions $(C.2)-(C.5)$ hold. Then the following holds:
\begin{align*}
\Big{|}\Big{|}n^{-1/2}\sum_{i=1}^{n}\epsilon_i\bm{X}_i\Big{|}\Big{|}=o(\log n), \;\;\text{w.p.}\; 1 .
\end{align*}
\end{lem}

Proof of Lemma \ref{lem:con}. Assume $s_n^2=\max_{j\in\{1,\dots,p\}}\sum_{i=1}^{n}\mathbf{E}X_{ij}^2\epsilon_i^2$. Note that by assumption (C.4), $s_n^2=O(n)$. Rest of the proof now follows in the same line as in the Lemma 4.1 of Chatterjee and Lahiri (2010).

\begin{lem}\label{lem:lassocon}
Suppose conditions $(C.1)-(C.5)$ hold. Then
\begin{align}\label{eqn:0a}
\big{|}\big{|}\bm{T}_n\big{|}\big{|}=O(\log n), \;\;\text{w.p.}\; 1 .
\end{align}
\end{lem}

Proof of Lemma \ref{lem:lassocon}. lemma \ref{lem:lassocon} can be proved using same line of arguments as in the proof of the Lemma 4.2 in Chatterjee and Lahiri (2010). We omit the details to save space. 

\begin{lem}\label{lem:moment}
Suppose conditions $(C.1)-(C.3)$ hold. Then the following is true:
\begin{align}\label{eqn:0b}
n^{-1}\Big{|}\Big{|}\sum_{i=1}^{n}\bm{X}_i\bm{X}_i^\prime \tilde{\epsilon}_i^2-\mathbf{E}\Big{(}\sum_{i=1}^{n}\bm{X}_i\bm{X}_i^\prime \epsilon_i^2\Big{)}\Big{|}\Big{|}+n^{-3/2}\sum_{i=1}^{n}||\bm{X}_i||^3|\tilde{\epsilon}_i|^3=o(1),\;\text{w.p.}\; 1.
\end{align}
\end{lem}

Proof of Lemma \ref{lem:moment}. First note that 
\begin{align}\label{eqn:a}
&n^{-1}\Big{|}\Big{|}\sum_{i=1}^{n}\bm{X}_i\bm{X}_i^\prime \tilde{\epsilon}_i^2-\mathbf{E}\Big{(}\sum_{i=1}^{n}\bm{X}_i\bm{X}_i^\prime \epsilon_i^2\Big{)}\Big{|}\Big{|}\nonumber \\
\leq& \Big{|}\Big{|}n^{-1}\sum_{i=1}^{n}\bm{X}_i\bm{X}_i^\prime \big(\tilde{\epsilon}_i^2-\epsilon_i^2\big{)}\Big{|}\Big{|}
+\Big{|}\Big{|}n^{-1}\sum_{i=1}^{n}\bm{X}_i\bm{X}_i^\prime \big(\epsilon_i^2-\mathbf{E}\epsilon_i^2\big{)}\Big{|}\Big{|} \nonumber\\ 
\leq& \sum_{j,k=1}^{p}\big{|}n^{-1}\sum_{i = 1}^{n}X_{ij}X_{ik}\big(\tilde{\epsilon}_i^2-\epsilon_i^2\big{)}\big{|}+ \sum_{j,k=1}^{p}\big{|}n^{-1}\sum_{i = 1}^{n}X_{ij}X_{ik}\big(\epsilon_i^2-\mathbf{E}\epsilon_i^2\big{)}\big{|}.
\end{align}

Again since $(\tilde{\epsilon}_i-\epsilon_i)^2=\big[\bm{X}_i^\prime\big(\bm{\beta}-\tilde{\bm{\beta}}_n\big)\big]\big[\bm{X}_i^\prime\big(\bm{\beta}-\tilde{\bm{\beta}}_n\big)+2\epsilon_i\big]$, hence it follows that 
\begin{align*}
\big{|}n^{-1}\sum_{i = 1}^{n}X_{ij}X_{ik}\big(\tilde{\epsilon}_i^2-\epsilon_i^2\big{)}\big{|}\leq& n^{-1}\Big(\sum_{i=1}^{n}||\bm{X}_i||^4\Big)||\tilde{\bm{\beta}}_n-\bm{\beta}||^2\\
&+2n^{-1}\Big(\sum_{i=1}^{n}||\bm{X}_i||^3|\epsilon_i|\Big)||\tilde{\bm{\beta}}_n-\bm{\beta}||.
\end{align*}

Now note that by Lemma \ref{lem:lassocon} $||\sqrt{n}(\tilde{\bm{\beta}}_n-\bm{\beta})||=O(\log n)\; w.p.\; 1$. Using this, Lemma \ref{lem:indcon} and the fact that 
\begin{equation}\label{eqn:b}
\Big[\max_{i\in \{1,\dots,n\}}||\bm{X}_i||\big(1+|\epsilon_i|\big)\Big]\leq \Big(\sum_{i=1}^{n}||\bm{X}_i||^2\big(1+|\epsilon_i|^2\big)\Big)^{1/2}=O(n^{1/2}), \;\;\;\;\;\;w.p.\; 1,
\end{equation}
we have
\begin{align*}
\big{|}n^{-1}\sum_{i = 1}^{n}X_{ij}X_{ik}\big(\tilde{\epsilon}_i^2-\epsilon_i^2\big{)}\big{|}=o(1), \;\;\;\;\;\;w.p.\; 1.
\end{align*}

On the other hand, following is the direct implication of Lemma \ref{lem:indcon}
\begin{align*}
\big{|}n^{-1}\sum_{i = 1}^{n}X_{ij}X_{ik}\big(\epsilon_i^2-\mathbf{E}\epsilon_i^2\big{)}\big{|}=o(1), \;\;\;\;\;\;w.p.\; 1.
\end{align*}

This shows that the first term on the LHS of \ref{eqn:a} is $o(1)$, $w.p.\; 1$. Next consider the second term on the LHS of \ref{eqn:a}. It is easy to see that
\begin{align}\label{eqn:e}
\sum_{i=1}^{n}||\bm{X}_i||^3\Big(|\tilde{\epsilon}_i|^3-|\epsilon_i|^3\Big)\leq& \Big(\sum_{i=1}^{n}||\bm{X}_i||^6\Big)||\tilde{\bm{\beta}}_n-\bm{\beta}||^3\nonumber\\
+&3\Big(\sum_{i=1}^{n}||\bm{X}_i||^5|\epsilon_i|\Big)||\tilde{\bm{\beta}}_n-\bm{\beta}||^2+3\Big(\sum_{i=1}^{n}||\bm{X}_i||^4|\epsilon_i|^2\Big)||\tilde{\bm{\beta}}_n-\bm{\beta}||
\end{align}

Using (\ref{eqn:b}), H\"older's inequality, Lemma \ref{lem:indcon} and Lemma \ref{lem:lassocon}, one can show from (\ref{eqn:e}) that

\begin{align*}
n^{-3/2}\sum_{i=1}^{n}||\bm{X}_i||^3\Big(|\tilde{\epsilon}_i|^3-|\epsilon_i|^3\Big)=o(1), \;\;\;\;\;\;w.p.\; 1
\end{align*}

Again it is easy to show that $n^{-3/2}\sum_{i=1}^{n}||\bm{X}_i||^3|\epsilon_i|^3=o(1)\; w.p.\; 1$.

Hence Lemma \ref{lem:moment} follows.

\begin{lem}\label{lem:BET}
Suppose conditions $(C.1)-(C.6)$ hold. Then 
\begin{align}\label{eqn:0c}
\mathcal{L}\Big{(}\mu_{G^*}^{-1}\bm{\tilde{W}}_n^{*}\Big{|}\mathscr{E}\Big{)}\xrightarrow{d} \mathbf{N}\big{(}\bm{0},\bm{\Sigma}\big{)}, \;\;\text{w.p.}\; 1,
\end{align}
\end{lem}

Proof of Lemma \ref{lem:BET}. 
Consider $A\in \mathscr{E}$ such that $\mathbf{P}(A)=1$ and on the the set $A$ the following holds:
\begin{align*}
n^{-1}\Big{|}\Big{|}\sum_{i=1}^{n}\bm{X}_i\bm{X}_i^\prime \tilde{\epsilon}_i^2-\mathbf{E}\Big{(}\sum_{i=1}^{n}\bm{X}_i\bm{X}_i^\prime \epsilon_i^2\Big{)}\Big{|}\Big{|}+n^{-3/2}\sum_{i=1}^{n}||\bm{X}_i||^3|\tilde{\epsilon}_i|^3=o(1).
\end{align*}

Hence, using Cramer-Wold device, it is enough to show that, on $A$,
\begin{align}\label{eqn:f}
\mathcal{L}\Big{(}\bm{t}^{\prime}\bm{\tilde{W}}_n^{*}\leq x\mu_{G^*}\Big{|}\mathscr{E}\Big{)}\xrightarrow{d} \mathbf{N}\big{(}\bm{0},\bm{t}^{\prime}\bm{\Sigma}\bm{t}\big{)},
\end{align}
for any $\bm{t}\in \mathcal{R}^p$ and $\bm{t}\neq \bm{0}$. Now due to Lemma \ref{lem:moment} and assumption (C.4), (\ref{eqn:f}) follows if we can show that on the set $A$ 
\begin{align*}
\sup_{x\in \mathcal{R}^p}\Big|\mathbf{P}_*\Big{(}\bm{t}^{\prime}\bm{\tilde{W}}_n^{*}\leq x\mu_{G^*}\Big)-\Phi\big(xs_n^{-1}(\bm{t})\big)\Big|=o(1)
\end{align*}
where $s_n^{2}(\bm{t})=\bm{t}^{\prime}\Big[n^{-1}\sum_{i=1}^{n}\bm{X}_i\bm{X}_i^\prime \tilde{\epsilon}_i^2\Big]\bm{t}$. Now due to Berry-Essen Theorem [cf. Bhattacharya and Rao  (1986)], on the set $A$
\begin{align*}
\sup_{x\in \mathcal{R}^p}&\Big|\mathbf{P}_*\Big{(}\bm{t}^{\prime}\bm{\tilde{W}}_n^{*}\leq x\mu_{G^*}\Big)-\Phi(xs_n^{-1}(\bm{t}))\Big|\\
& \leq (2.75)\dfrac{\sum_{i=1}^{n}\mathbf{E}_*\Big|n^{-1/2}\bm{t}^{\prime}\bm{X}_i\tilde{\epsilon}_i(G_i^*-\mu_{G^*})\Big|^3}{\mu_{G^*}^{3}\Big{\{}\bm{t}^{\prime}\Big[n^{-1}\sum_{i=1}^{n}\bm{X}_i\bm{X}_i^\prime \tilde{\epsilon}_i^2\Big]\bm{t}\Big{\}}^{3/2}}\\
& \leq (2.75)\dfrac{||\bm{t}||^3n^{-3/2}\sum_{i=1}^{n}||\bm{X}_i||^3|\tilde{\epsilon}_i|^3\mathbf{E}_*|G_i^*-\mu_{G^*}|^3}{\mu_{G^*}^{3}s_n^{3}(\bm{t})}\\
&=o(1).
\end{align*}
The last equality follows from Lemma \ref{lem:moment}. Therefore Lemma \ref{lem:BET} follows.\\

Proof of Theorem \ref{eqn:theo1}. The Perturbation bootstrap version of the Lasso estimator is given by
 \begin{align*}
\bm{\hat{\beta}_n^*} = \operatorname*{arg\,min}_{\bm{t}^*}&\Bigg[\sum_{i=1}^{n}(y_i - \bm{x}'_i \bm{t}^*)^2(G^*_i-\mu_{G^*}) \nonumber\\
&+\sum_{i=1}^{n}[\bm{x}'_i(\bm{t}^*-\tilde{\bm{\beta}}_n)]^2(2\mu_{G^*}-G_i^*)+\mu_{G^*}\lambda_n\sum_{j=1}^{p}|t_{j}^*|\Bigg].
\end{align*}

Now, writing $\bm{\hat{u}}_n^*=\sqrt{n}\big(\hat{\bm{\beta}}_n^*-\tilde{\bm{\beta}}_n\big)$, we have
 \begin{align}
\bm{\hat{u}}_n^{*}&= \operatorname*{arg\,min}_{\bm{v}^*}\Bigg[\bm{v}^{*\prime}\bm{C}_n\bm{v}^{*} -2\mu_{G^*}^{-1}\bm{v}^{*\prime}\bm{\tilde{W}}_n^{*}+\lambda_n\sum_{j=1}^{p}\Big(|\tilde{\beta}_{j,n}+\dfrac{v_{j}^*}{\sqrt{n}}|-|\tilde{\beta}_{j,n}|\Big)\Bigg]\nonumber\\\label{eqn:Zn1}
&= \operatorname*{arg\,min}_{\bm{v}^*} \bm{Z}_n^*(\bm{v}^*)\;\;\;\;\; \text{(say)}.
\end{align}
Define the set $B\in \mathscr{E}$ such that $\mathbf{P}(B)=1$ and on the the set $B$, (\ref{eqn:0a}), (\ref{eqn:0b}) and (\ref{eqn:0c}) hold. Note that due to the definition of $\tilde{\bm{\beta}}_n$, there exists $N(\omega)$ for each $\omega\in B$ such that for $n>N(\omega)$,
\begin{align*}
\bigg{\{} \begin{array}{ll} \tilde{\beta}_{j,n}=\hat{\beta}_{j,n}\;\;\; \text{and}\;\;\; sgn({\tilde{\beta}_{j,n}})=sgn({\beta_{j}})\;  \text{for}\;j\in \mathcal{A}\;\\ 
\tilde{\beta}_{j,n}=0\; \text{for}\;j\in \{1,\dots,p\}\setminus \mathcal{A}   \end{array} 
\end{align*}

Therefore, on the set $B$,
\begin{align*}
\bm{Z}_n^*(\bm{v})\xrightarrow{d} Z(\bm{v})
\end{align*}
where $Z(\bm{v})=\Bigg[\bm{v}^\prime\bm{C}\bm{v} -2\bm{v}^{\prime}\bm{W}+\lambda_0\Big(\sum_{j=1}^{p_0}sgn({{\beta}_{j}})v_j+\sum_{j=p_0+1}^{p}|v_j|\Big)\Bigg]$ and $\bm{W}$ follows $N(\bm{0},\bm{\Sigma})$ distribution. Hence it follows, by the results of Geyer(1996), that 
\begin{align*}
\sqrt{n}\big(\hat{\bm{\beta}}_n^*-\tilde{\bm{\beta}}_n\big)\xrightarrow{d} \operatorname*{arg\,min}_{\bm{v}} Z(\bm{v})
\end{align*}

Again it is shown in Lemma 3.1 of Wagener and Dette (2012) that 
\begin{align*}
\sqrt{n}\big(\hat{\bm{\beta}}_n-\bm{\beta}\big)\xrightarrow{d} \operatorname*{arg\,min}_{\bm{v}} Z(\bm{v})
\end{align*}

Therefore Theorem \ref{eqn:theo1} follows.

\vspace*{3mm}

Proof of Proposition \ref{prop:resi}.
Writing $\bm{\tilde{u}}_n^*=\sqrt{n}\big(\breve{\bm{\beta}}_n^*-\tilde{\bm{\beta}}_n\big)$, we have from (\ref{eq:relasso})
 \begin{align*}
\bm{\tilde{u}}_n^{*}= \operatorname*{arg\,min}_{\bm{v}}\Bigg[\bm{v}^{\prime}\bm{C}_n\bm{v} -2\bm{v}^{\prime}\Big{[}n^{-1/2}\sum_{i=1}^{n}\bm{X}_i r_i^*\Big{]}+\lambda_n\sum_{j=1}^{p}\Big(|\tilde{\beta}_{j,n}+\dfrac{v_{j}}{\sqrt{n}}|-|\tilde{\beta}_{j,n}|\Big)\Bigg]
\end{align*}

Note that $Var\Big{(}n^{-1/2}\sum_{i=1}^{n} \bm{X}_i r_i^*|\mathscr{E}\Big{)}=\Big{(}n^{-1}\sum_{i=1}^{n}\bm{X}_i \bm{X}_i^\prime\Big{)} \Big{(}n^{-1}\sum_{i=1}^{n}(\tilde{\epsilon}_i-\bar{\epsilon}_n)^2\Big{)}$ which converges to $s^2\bm{C}$ as $n\rightarrow \infty$. Therefore Proposition \ref{prop:resi} follows through the same line of arguments, as in Theorem \ref{eqn:theo1}.

\vspace*{3mm}

Proof of Theorem \ref{eqn:theo2}. Theorem \ref{eqn:theo2} follows by retracting the steps of Theorem \ref{eqn:theo1} and using Lemma 1 of Camponovo (2015).

\section{Conclusion} \label{sec:1.77}
Perturbation Bootstrap method is proposed in case of Lasso. It is shown that the perturbation bootstrap is capable of consistently approximating the distribution of Lasso in both the cases when the covariates are random and non-random and even when the errors are heteroscedastic. Residual bootstrap of Chattaerjee and Lahiri (2011) is shown to fail when the errors are heteroscedastic and covariates are non-random. Thus perturbation bootstrap is better than residual bootstrap in the sense of validity of bootstrap in heteroscedasticity. When the covariates are random then one can implement paired bootstrap, the validity of which is shown by Camponovo (2015). The implementation of paired bootstrap is significantly different from residual bootstrap. Thus one needs to select the residual and the paired bootstrap depending on the nature of the covariates, where as the perturbation bootstrap works irrespective of the nature of the covariates. Therefore the results in this paper establish the proposed perturbation bootstrap method as a more general bootstrap method than the resample-based bootstrap methods (residual and paired) as a tool of distributional approximation of Lasso.

\end{document}